\documentclass[letterpaper]{article} 
\usepackage{aaai2026}
\usepackage{times}  
\usepackage{helvet}  
\usepackage{courier}  
\usepackage[hyphens]{url}  
\usepackage{graphicx} 
\usepackage{natbib}  
\usepackage{caption} 
\usepackage{algorithm}
\usepackage{algorithmic}
\usepackage{booktabs}
\usepackage{multirow}
\usepackage{pifont} 
\usepackage{amsmath}
\usepackage{amssymb}
\usepackage[most]{tcolorbox}

\usepackage{xcolor} 
\usepackage{graphicx}
\usepackage{subcaption}
\usepackage{newfloat}
\usepackage{listings}
\DeclareCaptionStyle{ruled}{labelfont=normalfont,labelsep=colon,strut=off} 
\floatstyle{ruled}
\newfloat{listing}{tb}{lst}{}
\floatname{listing}{Listing}
\nocopyright

\def\eg{\textit{e.g.,} }
\def\ie{\textit{i.e.,} }

\title{A System Model Generation Benchmark from Natural Language Requirements}

\author{
    Dongming Jin \textsuperscript{\rm 1,2},
    Zhi Jin \textsuperscript{\rm 1,2}\thanks{Corresponding author}, 
    Linyu Li \textsuperscript{\rm 1,2}, 
    Zheng Fang \textsuperscript{\rm 1,2}, 
    Jia Li \textsuperscript{\rm 3}, 
    Xiaohong Chen \textsuperscript{\rm 4},
    Yixing Luo \textsuperscript{\rm 5}
}
\affiliations{
    \textsuperscript{\rm 1}Key Laboratory of High Confidence Software Technologies (Peking University), Ministry of Education, Beijing, China\\
    \textsuperscript{\rm 2}School of Computer Science, Peking University, Beijing, China \\
    \textsuperscript{\rm 3}Wuhan University, Wuhan, China \\
    \textsuperscript{\rm 4}East China Normal University, Shanghai, China \\
    \textsuperscript{\rm 5}Beijing Institute of Control Engineering, Beijing, China \\
}

\usepackage{bibentry}

\begin{document}

\maketitle

\begin{abstract}

System models, a critical artifact in software development, provide a formal abstraction of both the structural and behavioral aspects of software systems, 
which can facilitate the early requirements analysis and architecture design. However, developing system models remains challenging due to the specific syntax of model description languages and the relative scarcity of public model examples. While large language models (LLMs) have shown promise in generating code with programming languages and could potentially aid in system model development, no benchmarks currently exist for evaluating their ability to generate system models with specific description languages. We present SysMBench, which comprises 151 human-curated scenarios spanning a wide range of popular domains and varying difficulty levels. Each scenario mainly comprises a natural language requirements description, a system model expressed in a specific model description language, and a visualized system model diagram. The requirements description is fed as user input to the LLM, the system model with description language is used to verify if the generated system model conforms to the requirements, and the visualized diagram serves to support manual validation. We introduce SysMEval, a semantic-aware evaluation metric to evaluate the quality of generated system models. We evaluate 17 popular LLMs on this task with three traditional metrics and SysMEval, from directly prompting to three commonly used enhancement strategies. Our in-depth evaluation shows that LLMs perform poorly on SysMBench, with the highest BLEU of 4\% and SysMEval-F1 of 62\%. We release the SysMBench and its evaluation framework to enable future research on LLM-based system model generation.

\end{abstract}


\section{Introduction} \label{sec:intro}

\begin{figure}
    \centering
    \includegraphics[width=0.99\linewidth]{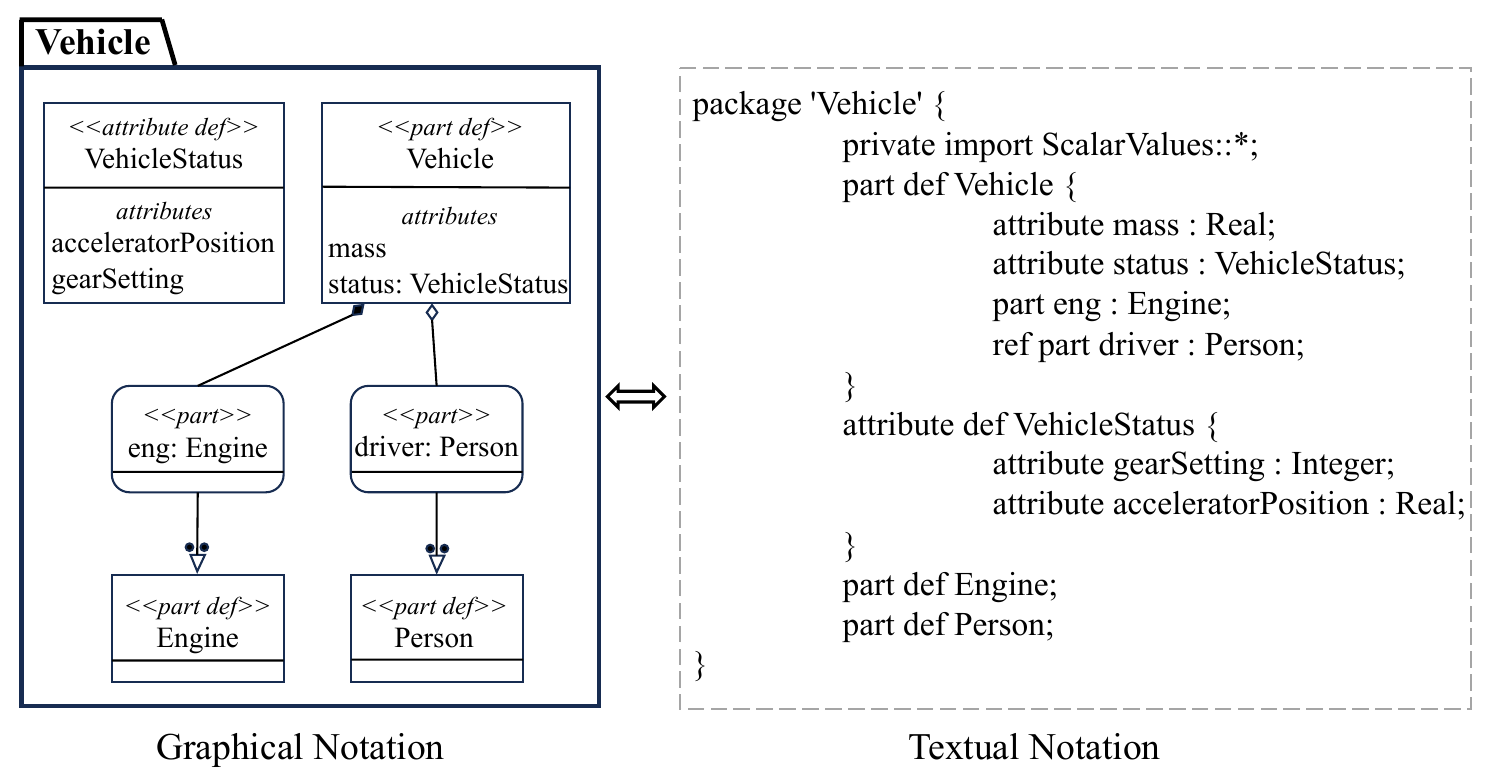}
    \caption{A system model example with graphical notation and textual notation}
    \label{fig:teaser}
\end{figure}

System models are widely recognized as critical artifacts in the software development process~\cite{basha2012model}, particularly for large-scale and safety-critical systems~\cite{ahlbrecht2024exploring}. As illustrated in Figure~\ref{fig:teaser}, they provide structured representations of a system's architecture and behaviors, thereby facilitating rigorous requirements analysis and architecture design. Many organizations (\eg NASA and Airbus) have utilized system models to reduce late-stage defects and accelerate development cycles. NASA reported that the use of system models halved the preliminary design review cycle for Orion spacecraft's electrical architecture, reducing it from six weeks to three weeks~\cite{lindsey2020reliability}. Against this backdrop, constructing high-quality system models from natural language requirements has become an increasingly important concern in the software engineering community.

Developing system models remains a non-trivial task. For one, it requires not only a deep understanding of the underlying system requirements, but also proficiency in the specific syntax and semantics of model description language such as SysML~\cite{hause2006sysml} or AADL~\cite{feiler2012model}. These languages are often complex, verbose, and domain-specific, making them difficult to master, particularly for practitioners without formal training in model-driven engineering. Moreover, publicly available system model examples are scarce, limiting the opportunities for data-driven learning or tool development. As a result, the manual construction of system models remains time-consuming and error-prone, posing a significant barrier to the widespread adoption of model-based practices in real-world projects.

Recent advances in large language models (LLMs) have demonstrated remarkable capabilities in translating natural language text into structured formal representations, such as source code~\cite{zhu2025domaineval}, SQL queries~\cite{cen2025sqlfixagent}, and formal expressions~\cite{cao2025informal}. Given these strengths, LLMs hold significant potential for assisting in system model development, particularly in translating natural language requirements into system models written in model description language. Unfortunately, the ability of existing LLMs to effectively generate system models remains uncertain, since there exist no systematic studies or datasets/benchmarks available to quantitatively evaluate LLM performance on system model generation.

To bridge this gap, we introduce SysMBench, the first benchmark designed to evaluate the capability of LLMs in generating system models with model description language from natural language requirements. SysMBench's dataset includes 151 human-curated scenarios (comparable in size to the HumanEval dataset~\cite{liu2023your} which contains 164 programming problems) covering a wide range of popular domains, ranging from simple to highly challenging scenarios that contain hundreds of lines of system model descriptions (LoM). Each scenario mainly contains three components: (1) a natural language requirements description (\eg ``Define the basic information of vehicles'') fed as user input to the LLMs; (2) a reference system model written in a specific modeling language (\ie SysML) serving as ground truth to check against the LLM-generated models; and (3) a visualized diagram of the system model to support human validation and qualitative analysis. This benchmark enables both automatic and manual assessment of generated models, laying a foundation for systematic evaluation and future advancement in LLM-based system model generation.

To address the limitation of traditional metrics for this task, which primarily measure surface-level similarity, we introduce a semantics-aware evaluation metric named SysMEval. Instead of relying on surface-level string comparison, SysMEval decomposes each candidate model into a set of atomic semantic claims. Each claim represents a minimal, verifiable statement about structural or behavioral elements, \eg an \textit{attribute} or a \textit{port connection}. Levering the reasoning capabilities of GPT-4~\cite{achiam2023gpt}, SysMEval checks whether each atomic claim is explicitly supported by the reference system model. The final score is computed as the proportion of supported claims among all extracted claims, offering a fine-grained assessment of both correctness and completeness. By aligning evaluation with the semantics of system models, SysMEval provides a more reliable and informative metric for assessing the quality of the generated system models.

Finally, we conduct a comprehensive evaluation of 17 popular LLMs on the SysMBench, covering 10 general LLMs and 7 code LLMs. To simulate realistic usage scenarios, we assess each LLM under multiple widely used prompting settings, including zero-shot, few-shot, chain-of-thought, and grammar prompting. We reported results across multiple traditional metrics (\ie BLEU, ROUGE-L, and BertScore) and our proposed SysMEval to capture both surface-level and semantics-level generation quality. Our empirical findings reveal that all LLMs struggle significantly with the system model generation task. Even the best-performing LLM (\ie \textit{Qwen3-32B}) achieves only 62\% on SysM-F1 and 2.4\% on BLUE, indicating the limitations in current LLMs' ability to generate a system model based on a given natural language requirements description.

In summary, this paper makes the following contributions.

\begin{itemize}
    \item We introduce the first system model generation benchmark from natural language requirements named SysMBench, comprising 151 scenarios across multiple domains and difficulty levels.
    \item We propose a semantics-aware evaluation metric named SysMEval, which assesses the correctness and completeness of generated models based on structured claim comparison using LLMs.
    \item We conduct a thorough evaluation of 17 mainstream LLMs using four metrics and four prompting strategies, revealing the current limitations of LLMs in this challenging task.
\end{itemize}
\section{Related Work} \label{sec:related}

\subsection{System Model Generation Benchmark.}

Benchmarks for structured language generation are well established in programming languages (\eg Python~\cite{austin2021program} and SQL~\cite{yu2018spider}). However, there is no comparable, standardized benchmark for generating system models from natural language requirements in declarative modeling languages such as SysML. Prior work~\cite{jin2024evaluation} typically reformulates the task as information extraction of model elements (\ie various entities and relations) and releases only small, hand-crafted datasets (often fewer than ten scenarios), which lack scale and domain diversity and do not assess end-to-end generation under a model description language. Unlike imperative programming languages, system modeling languages are declarative. While related benchmark exist for other declarative formalisms (\eg VHDL-Eval~\cite{vijayaraghavan2024vhdl} for hardware description and LTL-pattern~\cite{hahn2022formal} for temporal logic), they target domain-specific specifications rather than SysML-based system modeling. This motivates a benchmark grounded in a specific modeling language. To fill this gap, we introduce SysMBench, the first public benchmark for SysML-based system model generation from natural-language requirements.

\begin{figure*}
    \centering
    \includegraphics[width=0.99\linewidth]{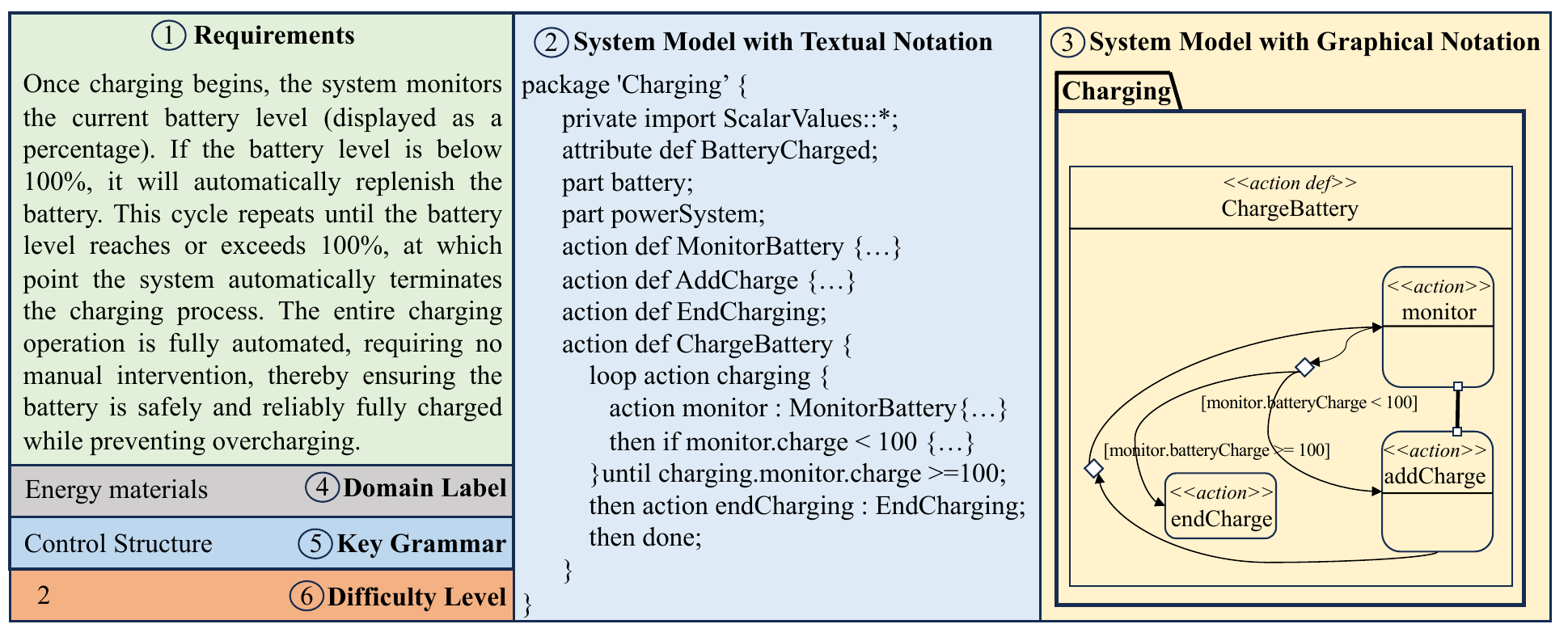}
    \caption{An overview of SysMBench. Each sample consists of six components.}
    \label{fig:overview}
\end{figure*}

\subsection{System Model Evaluation Metrics.} Evaluating system models expressed in a specific modeling language poses challenges distinct from free-form text generation. Text similarity metrics (\eg BLUE~\cite{papineni2002bleu} and ROUGE~\cite{lin-2004-rouge}) and human evaluation do not work well for evaluating system models with a specific modeling language for various reasons. Text similarity metrics and embedding-based scores (\eg bertscore~\cite{zhang2019bertscore}) are convenient but operate primarily on strings and thus fail to capture syntactic correctness, structural correctness, and deeper semantics. Even minor dissimilarities can result in significant problems. Human expert evaluation~\cite{li2023collaborative} is more reliable but costly, slow, and difficult to scale. Structure-aware metrics (\eg incorporating abstract syntax trees~\cite{ren2020codebleu}) better assess structural plausibility yet still struggle to judge semantic equivalence and constraint preservation. To address these limitations, we propose SysMEval, which decomposes both reference and predicted models into atomic semantic assertions and computes assertion-level precision and recall to determine correctness and completeness.

\section{SysMBench} \label{sec:bench}


\subsection{Overview}
SysMBench aims to facilitate the development and evaluation of automated system model generation techniques. As shown in Figure~\ref{fig:overview}, each sample in SysMBench consists of six components. \ding{182} \textbf{Requirements:} An English natural language description detailing the functional requirements of the software system. \ding{183} \textbf{System Model with Textual Notation:} A developer-written textual implementation of the system model using SysML syntax. \ding{184} \textbf{System Model with Graphical Notation:} A corresponding visualized system model diagram automatically rendered from the textual representation. \ding{185} \textbf{Domain Label:} The application domain of the system. \ding{186} \textbf{Key Grammar:} A key grammar to construct the system model. \ding{187} \textbf{Difficulty Level:} A assigned difficulty level ranging from 1 to 5.

\subsection{Benchmark Construction Pipeline}

The construction pipeline consists of five stages as follows.

\textbf{Stage 1: System Model Collection and Preprocessing.} We crawl system models with textual notation from publicly available SysML teaching materials and example repositories~\cite{sysml-v2}, resulting in a list of 161 unique system models across multiple domains. However, due to their origin from teaching materials, some system models suffer from three types of issues. \textit{(1) Lack practical scenarios:} the system models are only used to demonstrate some grammar usage (\eg comment) and do not contain actual scenarios. \textit{(2) Cross reference:} some model reference elements are defined in external files. \textit{(3) The name lacks semantics:} some system models are poorly named and do not reflect their functionality, \eg ``example1''. The examples of these three issues is provided in Appendix A. To address these issues, the first author manually reviewed and preprocessed all models, \ie the models without practical scenarios were removed, referenced elements from other files were manually inlined to make each system model self-contained, and all model names were revised to accurately reflect their semantics. This preprocessing ensures that each system model is independent, meaningful, and suitable for downstream generation and evaluation.

\textbf{Stage 2: Natural Language Requirements Annotation.} We assembled an annotation team consisting of two graduate-level software engineering students from a prestigious university. Each annotator has over four years of experience in software engineering and is familiar with system modeling concepts. The annotators are not co-authors and obtain adequate payments. To ensure quality, we first establish two key criteria for requirements through discussions with annotators. \textit{(1) Naturalness:} ensures the requirements read like a natural description from the perspective of a real-world stakeholder. \textit{(2) Completeness:} all structural and behavioral aspects of a system model must be reflected in the natural language requirements. During the annotation process, each requirement undergoes a dual-annotation process, with one annotator assigned to its initial drafting and another responsible for meticulous double-checking.

\begin{table*}[]
\centering
\begin{tabular}{lccccccc}
\toprule
\multirow{2}{*}{\textbf{Domain}} & 
\multirow{2}{*}{\textbf{Number}} & 
\multicolumn{3}{c}{\textbf{Requirement Tokens}} & 
\multicolumn{3}{c}{\textbf{System Model Lines}} \\
                                 &                                  & \textbf{Avg}  & \textbf{Max}  & \textbf{Min}  & \textbf{Avg}  & \textbf{Max} & \textbf{Min} \\ \midrule
Vehicle Traffic                  & 108                                & 131          & 229             & 71              & 45           & 158            & 7              \\
Photography Technique            & 12                                 & 100          & 119             & 78              & 23           & 34             & 14             \\
Information Management           & 8                                  & 122          & 181             & 82              & 21           & 44             & 3              \\
Simulation Calculation           & 4                                  & 124          & 156             & 101             & 23           & 36             & 15             \\
Energy materials                 & 3                                  & 126          & 192             & 88              & 34           & 55             & 23             \\
Network Communication            & 3                                  & 125          & 130             & 118             & 30           & 34             & 24             \\
Fault diagnosis                  & 3                                  & 137          & 178             & 102             & 42           & 62             & 25             \\
Aerospace                        & 3                                  & 109          & 152             & 70              & 29           & 54             & 15             \\
Confidentiality and security     & 2                                  & 101          & 136             & 67              & 28           & 33             & 24             \\
Systems Engineering              & 2                                  & 153          & 196             & 110             & 52           & 73             & 32             \\
Embedded device                  & 1                                  & 85           & 85              & 85              & 16           & 16             & 16             \\
Medical Health                   & 1                                  & 135          & 135             & 135             & 144          & 144            & 144            \\
Water resource transportation    & 1                                  & 111          & 111             & 111             & 18           & 18             & 18             \\ \midrule
\textbf{Total}                   & \textbf{151}                       & \textbf{127}  & \textbf{229}     & \textbf{67}     & \textbf{41}  & \textbf{158}    & \textbf{3}    \\ \bottomrule
\end{tabular}
\caption{Statistics of our SysMBench.}
\label{tab:data}
\end{table*}

\textbf{Stage 3: Requirements and Model Validation.} To ensure the quality of our SysMBench, we conduct model-side and requirements-side validation. \textit{(1) System Model Validation.} Since some system models were modified during preprocessing, the first author manually loaded each system model with textual notation into a SysML interpreter to ensure it compiles correctly into a valid system model diagram with graphical notation. \textit{(2) Requirements Validation.} For each requirements description, we verify that all elements in the system model are traceable to at least one sentence, and the description does not contain extra content. 

\textbf{Stage 4: Domain Label Annotation.} To annotate each sample's domain label, we manually design a domain taxonomy. Specifically, we read the office tutorial from the SysML organization. Based on this tutorial, we determine the top 10 domains that frequently occur. Finally, we invited the annotation team to annotate domain labels for each sample based on the requirements and our taxonomy. During the domain labeling process, the annotators found that five samples did not belong to any domain in our taxonomy. Thus, we add three additional domains to solve this issue. The domain taxonomy can be found in Appendix B.

\textbf{Stage 5: Key Grammar Label Annotation.} This label indicates that generating a system model requires mastering key grammar. Due to our system model from teaching materials, all system models have been classified into a set of categories by their official developers, which indicates that a system model is used to demonstrate which key grammar. Thus, we take the categories as our grammar taxonomy and the classification of each sample as its key grammar label. The grammar taxonomy and the grammar label distribution can be found in Appendix C.

\textbf{Stage 5: Difficulty Level Creation.} We introduce a system of difficulty levels for system model generation problems. We recognize that determining these levels is inherently ambiguous and subjective, akin to the informal designations used by online programming platforms (\eg LeetCode~\cite{leetcode}) and in existing research~\cite{luo2023wizardcoder}. Nevertheless, we propose an approximation that can calculate a difficulty level by parsing the system model expressed in a specific model description language, which is based on the LoM. The rules to label difficulty level and the difficulty level distribution can be found in Appendix D.

\subsection{Benchmark Statistics}

Table~\ref{tab:data} summarizes the core statistics of SysMBench, which contains 151 samples across 13 domains. The domain distribution is naturally uneven (\ie vehicle traffic domain contribute many samples) due to the varying availability of public teaching materials. To mitigate this bias, we report both macro-averaged in our experiments. In terms of the size, requirement description contains about 127 tokens on average, ranging from 67 to 229. The corresponding system models have an average of 41 lines in textual notation, with a minimum of 3 lines and a maximum of 158 lines. 
This reflects the diversity of real-world modeling tasks covered by SysMBench.



\section{SysMEval Metrics} \label{sec:stats}

We design a novel semantics-aware evaluation metric named \textbf{SysMEval} to assess the quality of automatically generated system models. 



\subsection{Key Ideas}

Inspired by Fact-QA~\cite{fernandez-etal-2024-syllabusqa} and FActScore~\cite{min-etal-2023-factscore}, SysMEval is based on two key ideas.

\textbf{Key idea 1: Atomic component or behavior as a unit.} A system model is composed of multiple granular elements that define either structural components (\eg \textit{part} and \textit{attribute}) or behavioral aspects (\eg \textit{action} and \textit{control structure}). SysMEval treats each such element as an atomic modeling claim and assigns them an equal weight of
importance, enabling fine-grained evaluation that aligns with the semantics of system-level design.

\begin{table*}[]
    \centering
    \begin{tabular}{lllcccccc}
\toprule
\multicolumn{3}{c}{\textbf{LLMs}}                                 & \multicolumn{6}{c}{\textbf{Evaluation Metric (\%)} } \\
\textbf{Family} & \textbf{Name}                   & \textbf{Size} & \multicolumn{1}{l}{\textbf{BLEU}} & \multicolumn{1}{l}{\textbf{ROGUE}} & \multicolumn{1}{l}{\textbf{BertScore}} & \multicolumn{1}{l}{\textbf{SysM-P}} & \textbf{SysM-R} &\textbf{SysM-F1}\\ \midrule
\multicolumn{9}{c}{\textbf{General LLMs}} \\ \midrule
GPT-4           & gpt-4.1-2025-04-14              & ?             &                                   2.0&                                    41&                                     65&                                         \textbf{71}& 46&56*\\
Claude 3        & Claude 3 Opus                   & ?             &                                   2.6&                                    42&                                     62&                                         70& 56&59*\\
DeepSeek        & DeepSeek R1                     & 685B          &                                   \textbf{4.0}&                                    \textbf{47}&                                     \textbf{69}&                                         58& \textbf{60}&53\\
Mistral         & Mistral-7B-instruct& 7B           &                                   1.0&                                    42&                                     59&                                         45& 48&46\\
Qwen3           & Qwen3-32B                   & 32B           &                                   2.4&                                    44&                                     66&                                         66& 58&\textbf{62*}\\
Gemma2& gemma-2-9b-it                  & 9B           &                                   1.2&                                    44&                                     59&                                         53& 59&56*\\
LLama3          & Llama-3.1-8B-Instruct           & 8B            &                                   1.0&                                    41&                                     57&                                         34& 35&34\\
InternLM        & internlm3-8b-instruct           & 8B            &                                   $<$0.1&                                    36&                                     52&                                         21& 47&29\\
Baichuan2       & Baichuan2-13B-Chat              & 13B           &                                   $<$0.1&                                    38&                                     52&                                         36& 58&45\\
ChatGLM3        & ChatGLM3-6B                     & 6B            &                                   $<$0.1&                                    37&                                     47&                                         17& 24&20\\ \midrule
\multicolumn{9}{c}{\textbf{Code LLMs}} \\ \midrule
DeepSeekCoder   & DeepSeek-Coder-V2-Lite      & 16B         &                                   0.5&                                    42&                                     58&                                         52& 63&57*\\
Mistral Open    & Codestral-22B-v0.1              & 22B           &                                   2.1&                                    45&                                     64&                                         48& 54&51\\
Phind           & Phind-CodeLlama-34B-v2          & 34B           &                                   1.5&                                    44&                                     62&                                         42& 54&48\\
Magicoder       & Magicoder-S-CL-7B               & 7B            &                                   0.1&                                    30&                                     60&                                         44& 46&45\\
WizardLM        & WizardCoder-15B-V1.0            & 15B           &                                   0.4&                                    42&                                     58&                                         49& 47&48\\
Code Llama      & CodeLlama-13b-Instruct-hf       & 13B           &                                   1.0&                                    31&                                     34&                                         31& 42&36\\
CodeGen2.5      & CodeGen2.5-7B-Instruct          & 7B            &                                   0.2&                                    36&                                     42&                                         35& 55&43\\ \bottomrule
\end{tabular}
    \caption{Performance on System Model Generation. \textbf{Bold} indicates the best result, and an asterisk (*) marks the top-5.}
    \label{tab:rq1}
\end{table*}

\textbf{Key idea 2: Reference-supported matching.} Unlike traditional metrics that rely on exact string or syntactic tree matching, SysMEval adopts a semantics-aware comparison strategy. An atomic modeling claim is correct if it is semantically supported by the reference model, even if the textual form or syntactic structure differs. This allows SysMEval to capture equivalence under renaming, reordering, or abstracted representations, which are common in real-world modeling practices.

\subsection{Definition}

Let $M_p$ denote the generated system model to be evaluated, and $M_t$ the ground-truth system model written by human experts. SysMEval decomposes both $M_p$ and $M_t$ into sets of atomic modeling claims, denoted as $S_p$ and $S_t$, respectively. The decomposition is performed via LLMs guided by a carefully designed prompt $P$. Based on the two sets, SysMEval defines two core metrics:

\textbf{SysMEval Precision (SysM-P):} the proportion of atomic claims in the generated system model ($S_p$) that are semantically supported by the reference system model ($S_t$).

\begin{equation}
    \text{SysMEval-P} = \frac{1}{|S_p|} \sum_{a \in S_p} \mathbb{I}[a \text{ is supported by } S_t]
\end{equation}

\textbf{SysMEval Recall (SysM-R):} the proportion of atomic claims in the reference system model ($S_t$) that are successfully recovered in the generated system model $S_p$.

\begin{equation}
    \text{SysMEval-R} = \frac{1}{|S_t|} \sum_{a \in S_t} \mathbb{I}[a \text{ is covered by } S_p]
\end{equation}

Based on the SysM-P and SysM-R, we can also calculate the F1 score as the \textbf{SysMEval-F1.}


\subsection{Implementation}

We implement SysMEval using GPT-4 (\ie \textit{gpt-4.1-2025-04-14}), which is responsible for two core tasks: (1) decomposing both the generated and reference system models into atomic modeling claim sets ($S_p$ and $S_t$), and (2) evaluating the semantic alignment between claims across the two sets. To guide the LLMs in performing these tasks, we adopt a chain-of-thought prompting strategy. The prompt $P$ for SysMEval and GPT-4 configuration details are provided in Appendix E.

\section{Experiments} \label{sec:evaluation}

\subsection{Base LLMs.}

We selected 17 popular LLMs from different families and evaluated them in SysMBench. They cover 10 general LLMs (\ie gpt-4.1-2025-04-14 and Qwen3-32B) and 7 Code LLMs (\ie starcoder2-15b, DeepSeek Coder-16B and CodeLLama-13B). The selected LLMs and their introduction can be found in Appendix F. We use official interfaces or implementations to reproduce these LLMs. We run these LLMs on 4 NVIDIA A6000-48G GPUs.

\begin{figure*}[t]
  \centering
  \begin{subfigure}{0.48\linewidth}
    \centering
    \includegraphics[width=\linewidth]{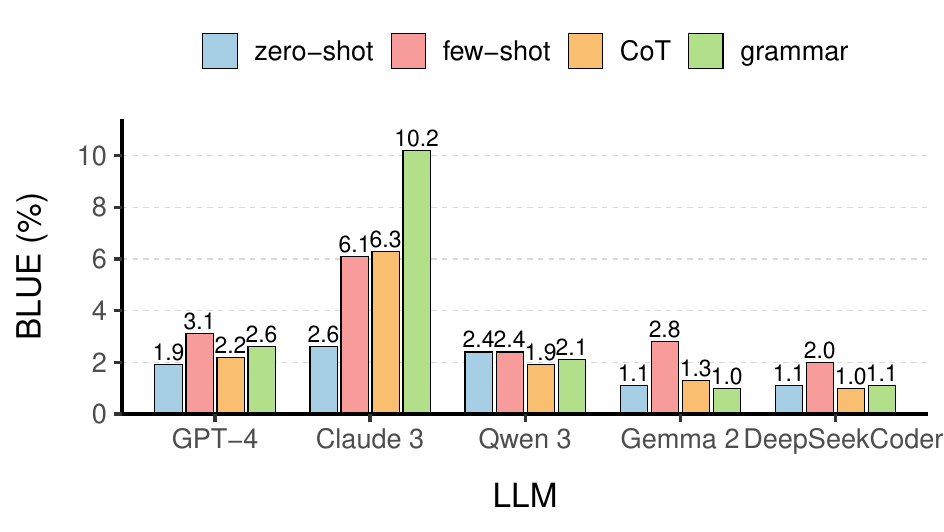}
    \label{fig:two-a}
  \end{subfigure}\hfill
  \begin{subfigure}{0.48\linewidth}
    \centering
    \includegraphics[width=\linewidth]{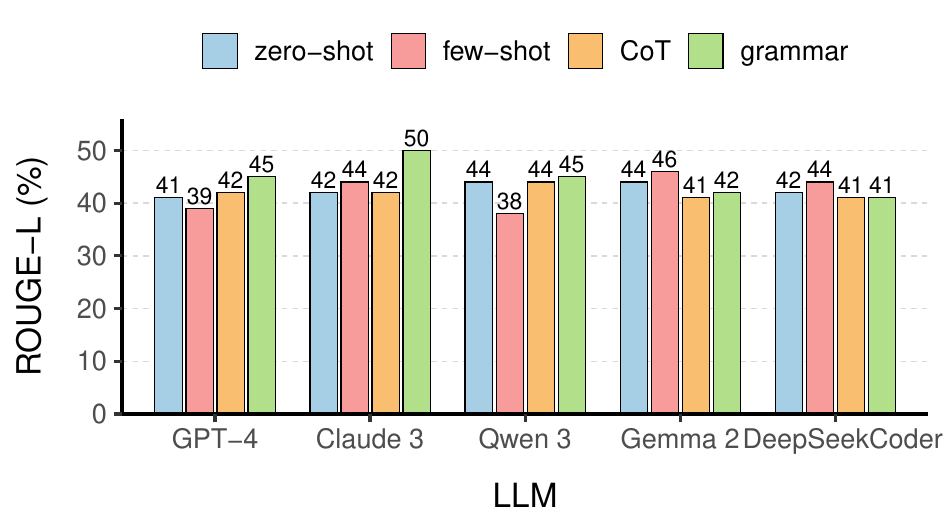}
    \label{fig:two-b}
  \end{subfigure}

  \medskip

  \begin{subfigure}{0.48\linewidth}
    \centering
    \includegraphics[width=\linewidth]{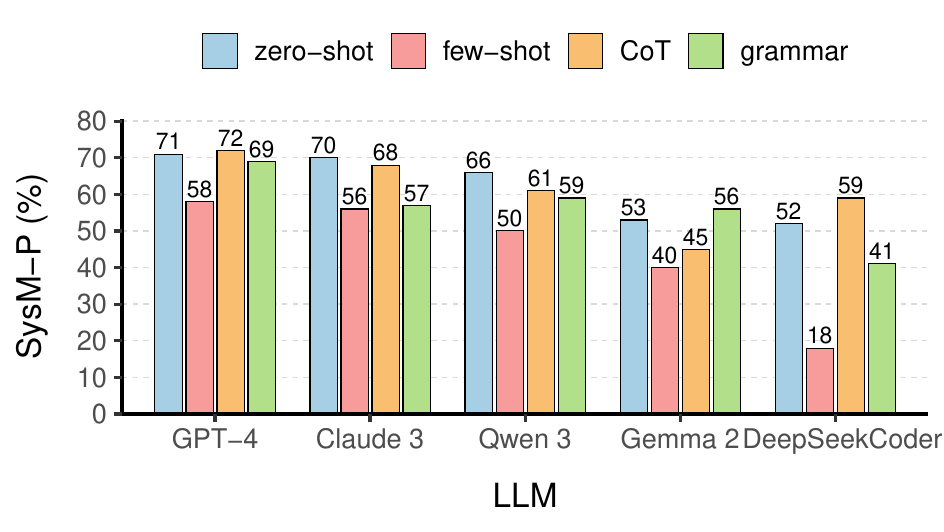}
    \label{fig:two-c}
  \end{subfigure}\hfill
  \begin{subfigure}{0.48\linewidth}
    \centering
    \includegraphics[width=\linewidth]{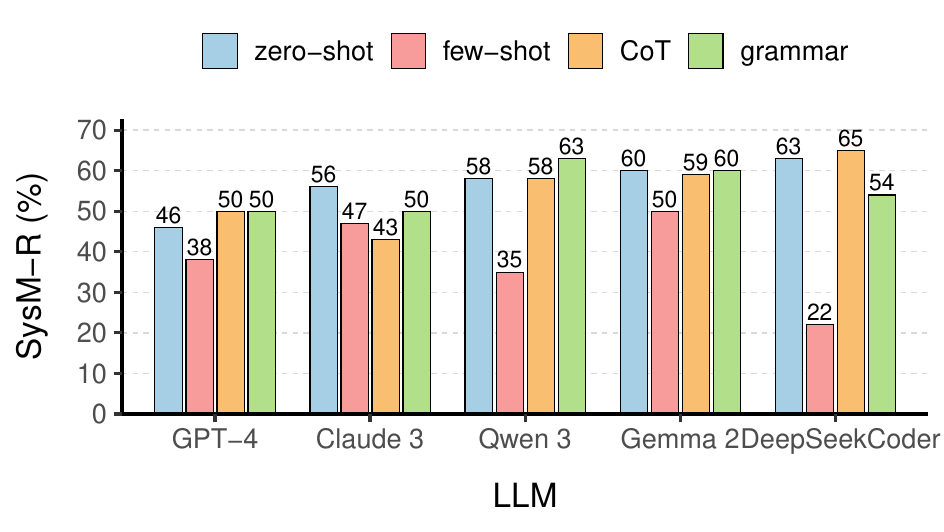}
    \label{fig:two-d}
  \end{subfigure}

  \caption{Average benchmark scores of top-5 LLMs when enhanced with differing strategies.}
  \label{fig:rq2}
\end{figure*}

\subsection{Selected Traditional Metrics.}

To comprehensively evaluate the LLMs from multiple perspectives, we also selected three widely adopted traditional metrics for domain-specific language generation. 

\textbf{BLEU.} The BLEU score~\cite{papineni2002bleu} is used to measure the token-level similarity between the generated system model and the ground truth. 

\textbf{ROUGE.} The ROUGE score~\cite{lin-2004-rouge} is used to quantify how much of the token overlap between the generated system model and the ground truth system model. Following common practice, ROUGE-L (F1) is selected to balance coverage and exactness.

\textbf{BertScore.} BERTScore~\cite{zhang2019bertscore} computes token-level semantic similarity using contextual embeddings from a pretrained encoder. We use \textit{bert-base-uncased} as the pretrained encoder and report the F1 variant of BERTScore.

\subsection{Evaluating LLMs Performance on SysMBench.}

We first evaluate the selected LLMs within a zero-shot prompting strategy to reflect their capability on our SysMBench. The prompt and experimental setting are provided in Appendix G. Table~\ref{tab:rq1} shows the evaluation results of different LLMs. We observe that DeepSeek R1 achieve best score in terms of BLEU, ROUGE, and BertScore, indicating that its peform better at both text and semantic levels. For our metrics, GPT-4, DeepSeek R1 and Qwen3-32B achieve the highest SysM-P, SysM-R, and SysM-F1 among all LLMs, respectively. However, all LLMs exhibit relatively low metrics, especially for the BLUE score.
For instance, DeepSeek R1 only achieves a BLUE score of 0.04 on our SysMBench. It suggests the difficulty of our dataset, and demonstrates that current LLMs are ineffective at generating system models with the SysML language. We also observe a clear SysM-P and SysM-R trade-off. GPT-4, DeepSeek R1, and Qwen 3-32B prioritize generating ``\textit{clean}'' system models with higher precision than recall (\ie SysM-P $>$ SysM-R), omitting long-tail elements. The other LLMs favor generating ``\textit{broad coverage}'' system model at the cost of precision. Finally, model size is not decisive for the quality of generated system models. Smaller and well-aligned LLMs (\eg Gemma-2-9b-it) can match or surpass much larger open-source LLMs.


\subsection{Assessing Common Enhancement Strategies.}
Prompt quality plays a pivotal role in an LLM's ability to generate accurate system models. To investigate whether practical prompt engineering techniques can narrow this sizeable performance gap observed in Table~\ref{tab:rq1} and reveal latent capabilities of these LLMs, we examine three widely adopted enhancement strategies (\ie few-shot, chain-of-thought, and grammar prompting). The specific prompt templates are provided in Appendix H. Figure~\ref{fig:rq2} shows the BLUE, ROUGE-L, SysM-P, and SysM-R scores achieved by the top-5 LLMs (ranked by SysM-F1 in Table~\ref{tab:rq1}) when applying each strategy. 


\textbf{Few-shot Prompting.} Few-shot prompting~\cite{brown2020language} involves providing a few examples to guide the LLMs. In our experiment, we just provide one example in the prompt. We observe that this technique consistently boosts surface-level text similarity, but diminishes semantic and structural accuracy. Specifically, BLEU rises slightly across most LLMs, whereas both SysM-P and SysM-R decline broadly. For example, Claude 3's BLEU rises from 2.6\% to 6.1\%, but its SysM-R falls from 56\% to 47\%. These results indicate that few-shot prompting can cause the LLMs to overfit to text-level patterns and narrow its coverage.  


\textbf{Chain-of-thought (CoT) Prompting.} Chain-of-thought prompting~\cite{wei2022chain} guides the LLMs through a step-by-step process to arrive at a solution, mimicking human logical progress. In our experiment, we ask LLMs to reason through extracting key elements and map them to the appropriate grammars in the SysML. Contrary to expectations, we find that this strategy rarely yields performance gains and can even degrade it. For instance, Claude 3's SysM-R score declines from 56\% to 43\%. Our error analysis (in Appendix I) reveals that the intermediate rationales frequently introduce redundant or partially formed constructs, which propagate into the final output and lower recall. Nevertheless, the explicit rationale produced by CoT remains valuable for post-doc system model debugging.


\begin{figure}
    \centering
    \includegraphics[width=0.99\linewidth]{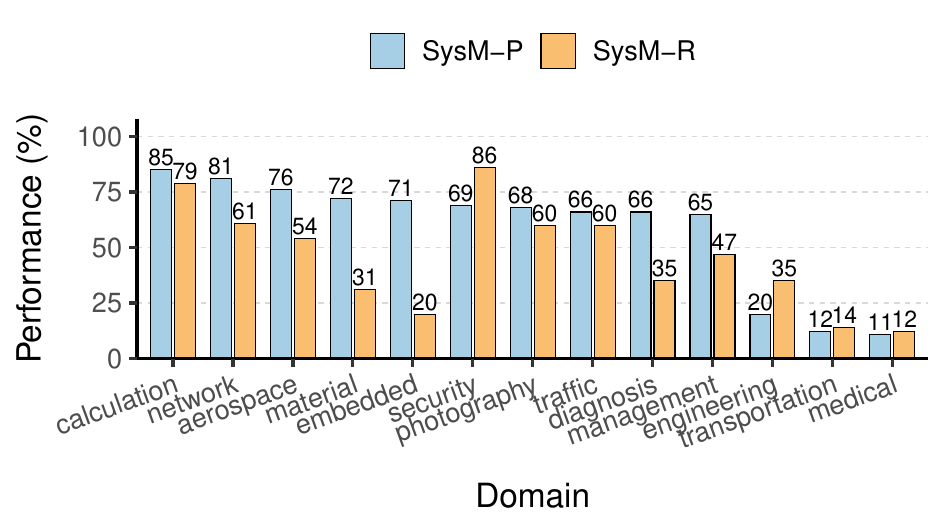}
    \caption{Performance in various application domains.}
    \label{fig:domain}
\end{figure}

\begin{table}[]
    \centering
    \begin{tabular}{cccccc}
\toprule
\multirow{2}{*}{Metrics} & \multicolumn{5}{c}{Difficulty Levels} \\ \cmidrule{2-6} 
                         & L1     & L2     & L3     & L4     & L5     \\ \midrule
BLUE                     & 2.8   & 2.4   & 2.3   & 0.4   & 0.5   \\
ROUGE                    & 45    & 46    & 43    & 36    & 40    \\
SysMEval-P               & 67    & 68    & 58    & 64    & 51    \\
SysMEval-R               & 52    & 62    & 61    & 76    & 51    \\ \bottomrule
\end{tabular}
    \caption{Performance in various difficulty levels.}
    \label{tab:difficult}
\end{table}

\textbf{Grammar Prompting.} Grammar prompting~\cite{wang2023grammar} injects an explicit Backus–Naur Form (BNF) specification for SysML directly into the prompt, steering the LLMs to follow the language's formal syntax and production rules. We observe that this grammar prompting can boost surface-level metrics that reward syntactic fidelity. For instance, Claude 3's BLEU score jumps from 2.6\% to 10.2\%. However, this improvement carries a semantic cost. Because the LLMs strives to satisfy every grammar rule, it tends to over-generate optional constructs, lowering semantic precision. For example, Claude 3's SysM-P drops from 70\% to 57\%. Thus, grammar prompting produces system models that look more ``complete'' on the surface but are less semantically accurate in practice.

\subsection{Performance across Different Domains and Levels.}
The domains and complexity determine the performance of LLMs on system model generation. To investigate how they affect the performance, we perform a fine-grained decomposition of the best-performing model in our benchmark (\ie Qwen 3-32B).

\textbf{Domain Results.} Figure~\ref{fig:domain} shows the SysMEval metrics of Qwen3-32B across different domains. We can observe that SysM-P is highest in \textit{calculation} and \textit{network}, with \textit{aerospace} close behind, whereas SysM-R peaks in \textit{security} and remains strong in \textit{calculation}. Several domains exhibit a large gap between SysM-P and SysM-R, such as \textit{materials}(72\% P vs. 31\%R). Thus, recall-oriented strategies (\eg explicit domain-knowledge infusion) are essential. Besides, safety-critical domains such as \textit{transportation} and \textit{medical} are much lower on both metrics, underscoring current limitations in high-reliability scenarios.

\textbf{Levels Results.} Table~\ref{tab:difficult} shows the results across different levels. We can observe that the surface metrics (\eg BLEU) show a monotonic decline from level 1 to level 5, whereas SysMEval is not monotonically. Interestingly, the precision–recall gap widens as the level grows. Balancing these competing objectives may be a crucial question for sustaining overall performance under complex scenarios.

\subsection{Key Grammar Analysis.}
\begin{figure}
    \centering
    \includegraphics[width=0.9\linewidth]{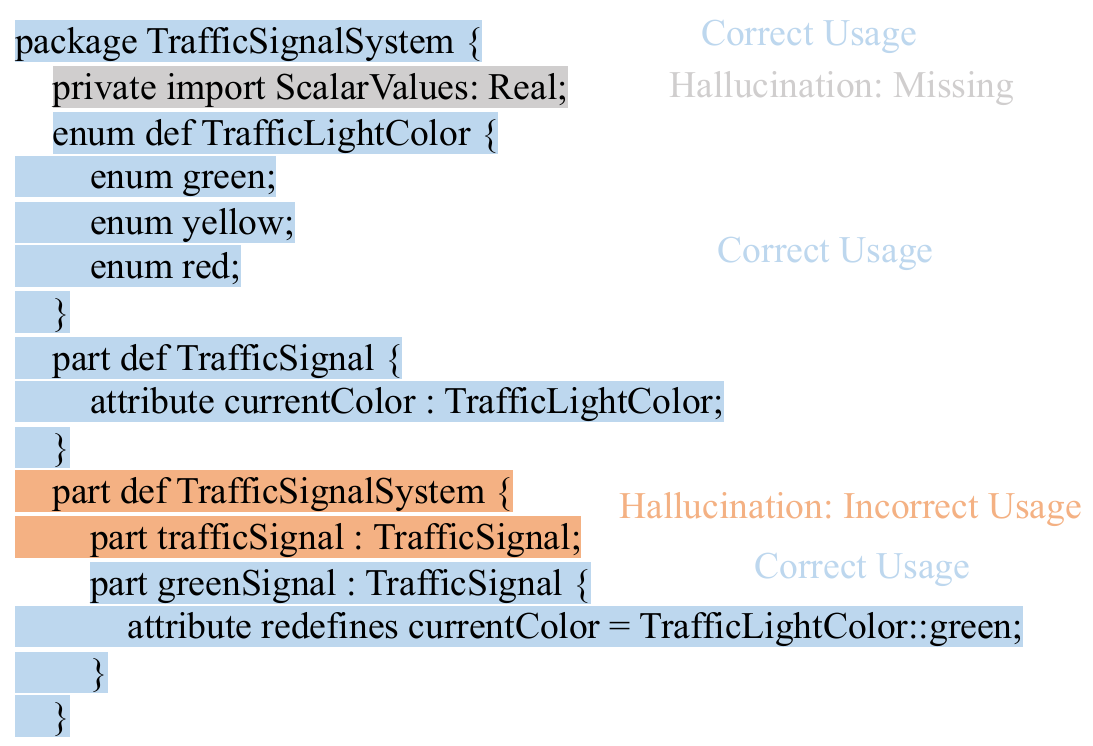}
    \caption{Incorrect system model generated by Claude 3.}
    \label{fig:case_study}
\end{figure}

To pinpoint the exact grammatical weaknesses that limit current LLMs, we manually inspect typical failure cases from SysMBench. Figure~\ref{fig:case_study} shows an incorrect system model generated by Claude 3 for the fifth sample (the ground-truth in Appendix J). While the model correctly applies the \textit{enumeration} and \textit{attribute} rules, two hallucination patterns emerge: \textbf{(1) Missing package import.} The model omits a required import for the \texttt{Real} type. \textbf{(2) Misusing specialization grammar.} A type-level specialization in the reference model is collapsed into an instance-level construct, and the \texttt{redefines} keyword is wrongly applied to that instance. More grammar-level analysis (\eg the performance across different grammars) is in Appendix K.

\section{Conclusion} \label{sec:conclusion}
We introduce SysMBench, the first dataset and benchmark capable of evaluating system model generation from natural language requirements by LLMs. SysMBench comprise 151 human-curated scenarios that cover diverse range of popular domains and varying difficulty levels. We also introduce SysMEval, a semantics-aware metrics for assess the correctness and completeness of system models. Our evaluation reveals that current LLMs, including GPT-4 perform poorly on SysMBench, with a SysM-F1 of 56\%. This underscores the need for advancements in LLM-based system model generation. We open-source SysMBench to facilitate future research in this domain.

\bibliography{aaai2026}

\section{Appendix}
\subsection{Table of Contents}
\begin{itemize}
    \item Appendix A: Issues in the Origin System Model.
    \item Appendix B: Domain Taxonomy for our SysMBench.
    \item Appendix C: Grammar Taxonomy and its Distribution.
    \item Appendix D: Difficulty and its Distribution.
    \item Appendix E: SysMEval Metrics Calculation.
    \item Appendix F: Studied LLMs.
    \item Appendix G: Experimental Setting for Evaluation.
    \item Appendix H: Prompt for Enhancement Strategies.
    \item Appendix I: Error Analysis.
    \item Appendix J. Ground Truth of Case Study.
    \item Appendix K. More Analysis on Key Grammar.
    
\end{itemize}

\subsection{A. Issues in the Origin System Model}

To illustrate the quality issues that often appear in publicly
available teaching materials, we select three representative
snippets as shown in Figure~\ref{fig:three_issues}\footnote{All examples are taken from the raw corpus before any cleaning or refactoring.}: 

\begin{itemize}
    \item \textbf{Lack practical scenarios.}  
    The first snippet (Figure~\ref{fig:issue_1}) is a self-contained \texttt{package} declaration whose sole purpose is to showcase \emph{import} and \emph{alias} syntax.  It does not correspond to any concrete use case or behavioural requirement, making it unsuitable for evaluating a model’s ability to capture real-world system semantics.
    \item \textbf{Cross reference.}  
    The second snippet (Figure~\ref{fig:issue_2}) references
    elements (\texttt{FuelOutPort}, \texttt{FuelInPort}, \texttt{FuelTankAssembly}, \texttt{Engine}) that are defined in external files.  When such cross-file dependencies are unavailable, the model becomes syntactically incomplete and cannot be parsed or analyzed in isolation.
    \item \textbf{Name lacks semantics.} The third snippet (Figure~\ref{fig:issue_3}) contains identifiers such as \texttt{Fuel}, \texttt{Person}, and \texttt{Vehicle}, but its enclosing package is simply called \texttt{Items Example}.  Non-descriptive names (e.g., \texttt{example1}, \texttt{demo}) obscure the intent of the model and hamper downstream tasks like retrieval and traceability recovery.
\end{itemize}

\begin{figure}[htbp]
    \centering
    \begin{subfigure}[t]{\columnwidth}
        \centering
        \begin{tcolorbox}[
            width=\columnwidth,
            colback=gray!4,
            colframe=gray!60,
            sharp corners,
            boxrule=0.4pt,
            listing only,
            listing options={style=sysml}
        ]
        package 'Package Example' \{ \\
        \hspace*{0.8em} public  import ISQ::TorqueValue; \\
        \hspace*{0.8em} private import ScalarValues::*; \\
        \hspace*{0.8em} private part def Automobile; \\
        \hspace*{0.8em} public  alias Car    for Automobile; \\
        \hspace*{0.8em} alias  Torque for ISQ::TorqueValue; \\
        \}
        \end{tcolorbox}
        \caption{\textit{Lack practical scenarios}}
        \label{fig:issue_1}
    \end{subfigure}

    \begin{subfigure}[t]{\columnwidth}
        \centering
        \begin{tcolorbox}[
            width=\columnwidth,
            colback=gray!4,
            colframe=gray!60,
            sharp corners,
            boxrule=0.4pt,
            listing only,
            listing options={style=sysml}
        ]
        package 'Interface Example' \{  \\
        \hspace*{0.8em} private import 'Port Example'::*; \\
        \hspace*{0.8em} part def Vehicle; \\
        \hspace*{0.8em} interface def FuelInterface \{ \\
        \hspace*{1.6em} end supplierPort : FuelOutPort; \\
        \hspace*{1.6em} end consumerPort : FuelInPort; \\
        \hspace*{0.8em} \} \\
        \hspace*{0.8em} part vehicle : Vehicle \{ \\
        \hspace*{1.6em} part tankAssy : FuelTankAssembly; \\
        \hspace*{1.6em} part eng : Engine; \\
        \hspace*{1.6em} interface : FuelInterface connect  \\
        \hspace*{2.4em} supplierPort ::> tankAssy.fuelTankPort to  \\
        \hspace*{2.4em} consumerPort ::> eng.engineFuelPort; \\
        \hspace*{0.8em} \} \\
        \}
        \end{tcolorbox}
        \caption{\textit{Cross reference}}
        \label{fig:issue_2}
    \end{subfigure}

    \begin{subfigure}[t]{\columnwidth}
        \centering
        \begin{tcolorbox}[
            width=\columnwidth,
            colback=gray!4,
            colframe=gray!60,
            sharp corners,
            boxrule=0.4pt,
            listing only,
            listing options={style=sysml}
        ]
        package 'Items Example' \{ \\
        \hspace*{0.8em} private import ScalarValues::*; \\
        \hspace*{0.8em} item def Fuel; \\
        \hspace*{0.8em} item def Person; \\
        \hspace*{0.8em} part def Vehicle \{ \\
        \hspace*{1.6em} attribute mass : Real; \\
        \hspace*{1.6em} ref item driver : Person; \\
        \hspace*{1.6em} part fuelTank \{ \\
        \hspace*{2.4em} item fuel: Fuel; \\
        \hspace*{1.6em} \} \\
        \hspace*{0.8em} \} \\
        \}
        \end{tcolorbox}
        \caption{\textit{The name lacks semantics}}
        \label{fig:issue_3}
    \end{subfigure}

    \caption{Illustrations of three common issues in SysML specification examples}
    \label{fig:three_issues}
\end{figure}

These examples motivate the data-cleaning heuristics, where we filter out grammar-only samples, resolve missing cross references, and rename packages and parts to reflect their functional roles.

\subsection{B. Domain Taxonomy for our SysMBench.}

Based on a careful review of the official SysML tutorial and the distribution of our 161 raw samples, we distilled ten core
application domains that appear most frequently.  During annotation, the team encountered five models that did not fit any of these ten domains, prompting the inclusion of three additional categories. Thus, the final taxonomy comprises \textbf{thirteen} domains, listed below with a brief description of the typical systems each domain covers.

\begin{itemize}
    \item \textbf{Vehicle Traffic:} Road\,/\,rail vehicles, traffic-control systems, and autonomous-driving components.
    \item \textbf{Photography Management:} Camera devices, photo-workflow management, and image-processing pipelines.
    \item \textbf{Information Management:} Databases, document repositories, and enterprise information systems.
    \item \textbf{Simulation Calculation:} Numerical simulators and high-performance computing for physics or engineering.
    \item \textbf{Energy Materials:} Battery packs, fuel cells, and energy-storage or material-processing subsystems.
    \item \textbf{Network Communication:} Wired/wireless network stacks, protocol suites, and router/switch designs.
    \item \textbf{Fault Diagnosis:} Health-monitoring or diagnostic frameworks for mechanical and electronic equipment.
    \item \textbf{Aerospace:} Aircraft, satellites, launch vehicles, and their on-board subsystems or ground support.
    \item \textbf{Confidentiality and Security:} Cryptographic modules, secure data flows, and access-control mechanisms.
    \item \textbf{System Engineering:} End-to-end system-of-systems architectures and generic SE process models.
    \item \textbf{Embedded Device:} Resource-constrained controllers, IoT nodes, and real-time firmware architectures.
    \item \textbf{Medical Health:} Diagnostic devices, patient-monitoring systems, and healthcare information flows.
    \item \textbf{Water Resource Transportation:} Water-distribution networks, pipeline control, and hydraulic transport.
\end{itemize}

The last three domains were introduced specifically to accommodate samples that could not be meaningfully mapped to the initial ten. For reproducibility, the full set of annotated labels is released together with SysMBench.

\subsection{C. Grammar Taxonomy and Its Distribution.}
To characterise the syntactic scope of \textsc{SysMBench}, we followed the official SysML v2 Textual Notation training material and distilled a fine-grained grammar taxonomy.  Each category below corresponds to a language construct that appears in at least one of the 161 benchmark models.  The same taxonomy was used by the annotation team to label every sample, enabling us to compute coverage statistics.

\begin{itemize}
    \item \textbf{Attribute:} Declares a property whose value describes a characteristic of a part, item, or package.
    \item \textbf{Generalization:} Specifies inheritance, allowing a definition to specialise and extend another.
    \item \textbf{Subsetting:} States that one feature’s value set is a subset of another feature within the same context.
    \item \textbf{Redefinition:} Overrides an inherited feature by changing its name, type, or multiplicity.
    \item \textbf{Enumeration:} Constrains an attribute to a predefined, discrete set of literal values.
    \item \textbf{Part:} A composable structural element that exists in space and time and may contain sub-parts.
    \item \textbf{Item:} A thing that can flow or be consumed/produced by the system (e.g.\ fuel, data); every part is an item, but not vice-versa.
    \item \textbf{Connection:} Establishes an explicit link between two features; if no custom definition is provided, the generic \texttt{Connection} is used.
    \item \textbf{Port:} An externally visible feature that exposes services or flows; a conjugate port (\texttt{\textasciitilde}) reverses direction for compatibility.
    \item \textbf{Function-based Behavior:} Captures behaviour in terms of actions, their parameters, and the flow/ succession that orders them.
    \item \textbf{Interface:} A reusable connection definition that binds compatible supplier and consumer ports.
    \item \textbf{State-based Behavior:} Describes discrete states, transitions, guards, and entry/exit behaviours.
    \item \textbf{Individual and Snapshot:} An \emph{individual} denotes exactly one instance; a \emph{snapshot} freezes that instance at a specific instant.
    \item \textbf{Binding Connector:} Asserts that the connected features share the same value within a given context.
    \item \textbf{Variant Configuration:} Language primitives for modelling product-line variability and selectable options.
    \item \textbf{Requirement:} A statement of expected capability or constraint that the system shall satisfy.
    \item \textbf{Verification:} Elements that define tests or analyses used to demonstrate requirement satisfaction.
    \item \textbf{Analysis and Trade:} Uses calculations and constraints to explore design alternatives and trade-offs.
    \item \textbf{View and Viewpoint:} A \emph{viewpoint} declares stakeholder concerns; a \emph{view} realises a viewpoint by filtering and presenting model elements.
    \item \textbf{Dependency:} A generic relationship expressing usage, ownership, or visibility between namespaces.
    \item \textbf{Model Constrainment:} Constraints—formal Boolean expressions—that must hold for the model to be valid.
    \item \textbf{Language Extension:} Embeds or references other languages (e.g.\ OCL, Alf) to enrich behavioural or analytical descriptions.
    \item \textbf{Expression:} A chain of feature references and operators that computes a value; also used in shorthand feature-value bindings.
    \item \textbf{SequenceModeling:} Represents temporal interaction using messages and event occurrences.
    \item \textbf{Flow Connection:} Transfers items or energy from an output feature to an input feature, optionally constrained by a flow definition.
    \item \textbf{Action Definition:} Declares a reusable behavioural primitive with typed parameters.
    \item \textbf{Action:} An invocation (usage) of an \emph{action definition} within a behaviour specification.
    \item \textbf{Conditional Succession:} Orders actions with a guard that determines whether the successor executes.
    \item \textbf{Control Structure:} Higher-level constructs such as \texttt{loop}, \texttt{if}, and \texttt{until} that orchestrate actions.
    \item \textbf{Assignment Action:} Sets the value of a property or parameter during behaviour execution.
    \item \textbf{Message:} A discrete piece of communication sent between behavioural entities in sequence modelling.
    \item \textbf{Opaque Action:} An action whose internal behaviour is defined in an external language or left unspecified.
    \item \textbf{State Definition:} Declares the set of allowable states and their sub-state hierarchy.
    \item \textbf{State:} A concrete condition during which specific invariants hold and activities may execute.
    \item \textbf{Transition:} A directed relationship that moves execution from one state to another when its trigger and guard are satisfied.
    \item \textbf{Occurrence:} A general term for an event or condition that happens at a point in time.
    \item \textbf{Individual:} (Repeated for completeness) A unique instance distinguished from all others of its type.
    \item \textbf{Calculation:} A reusable computation that returns a value, often referenced by constraints or analyses.
    \item \textbf{Constraint:} A Boolean expression that restricts values or relationships within the model.
    \item \textbf{Analysis:} A structured evaluation, often quantitative, performed on the model or its parameters.
    \item \textbf{Use Case:} Captures an external actor’s goal and the system behaviour required to achieve it.
    \item \textbf{Variability:} (Alias for \emph{Variant Configuration}) Explicit modelling of optional or alternative features.
    \item \textbf{Functional Allocation:} Maps behaviours (functions) onto structural elements (parts or items).
    \item \textbf{Metadata:} Ancillary information—such as author, version, or tags—attached to model elements.
    \item \textbf{Filtering:} A view mechanism that selects a subset of model elements based on criteria.
    \item \textbf{View:} (Alias listed for clarity) A concrete projection of the model that conforms to a viewpoint.
    \item \textbf{Package:} A namespace that organises definitions and controls their visibility and import.
\end{itemize}

\subsection{D. Difficulty and Its Distribution.}
We approximate the difficulty of a system model generation task by the Lines of Model (LoM) contained in its SysML text file, which is the number of non-blank, non-comment lines after preprocessing.   Although difficulty is inherently subjective, prior work on program synthesis benchmarks has demonstrated that source-length provides a practical first-order signal (e.g., LeetCode’s ''easy/medium/hard'' tags or the code length strata. We therefore partition the 151 samples into five levels via empirically chosen LoM break-points.

Table~\ref{tab:difficulty_distribution} summarises the resulting frequency of samples in each difficulty tier. Over 79\% of the corpus falls within Levels 1–2, indicating that most textbook SysML examples remain succinct.  Nevertheless, we retain a modest number of large-scale models (Levels 4–5) to ensure that \textsc{SysMBench} also assesses an LLM’s ability to handle longer, more intricate inputs.

\begin{table}[htbp]
    \centering
    \begin{tabular}{ccc}
        \toprule
        \textbf{Difficulty Level} & \multicolumn{1}{c}{\textbf{LoM Range}} & \textbf{\#Samples} \\
        \midrule
        1 & \(\text{LoM} < 30\)    & 63 \\
        2 & \(30 \le \text{LoM} < 60\)  & 64 \\
        3 & \(60 \le \text{LoM} < 90\)  & 12 \\
        4 & \(90 \le \text{LoM} < 120\) & 6  \\
        5 & \(\text{LoM} \ge 120\) & 6  \\
        \bottomrule
    \end{tabular}
    \caption{Difficulty-level distribution of the \textsc{SysMBench} corpus.}
    \label{tab:difficulty_distribution}
\end{table}

\subsection{E. SysMEval Metric Calculation.}

SysMEval quantifies how closely a generated SysML model
matches a reference model along two complementary axes:
\emph{precision} (SysM-P) and \emph{recall} (SysM-R).
Given a pair \(\langle \text{reference}, \text{generated} \rangle\), GPT-4 is prompted to

\begin{enumerate}
  \item decompose each model into a set of \emph{atomic modelling claims}
  \item align the two claim sets semantically, ignoring purely syntactic differences 
  \item return a score in the form Score: matches/total.
\end{enumerate}

\noindent

Figure~\ref{fig:sysm-p} and Figure~\ref{fig:sysm-r} list the complete user prompts for SysM-P and SysM-R, respectively.

\begin{figure}[htbp]
    \centering
    \begin{tcolorbox}[
        width=\columnwidth,            
        colback=gray!4,                
        colframe=gray!60,              
        sharp corners,                 
        boxrule=0.4pt,                 
        listing only,                  
        listing options={style=sysml}  
    ]
    Your task is to evaluate the precision of a generated system model. You will be given a reference system model and a generated system model. Please perform the following steps: \\
    \\
    1. List all atomic modeling claims made by the generated system model. Each atomic claim should correspond to a minimal, meaningful modeling element (e.g., the definition of a part, the declaration of an attribute, the use of types, or structural relations like containment or reference). \\
    2. For each atomic claim in the generated model, determine whether it is supported by the reference model (i.e., the reference model contains the same or equivalent element). \\
    3. Summarize the results using the format: Score: number of supported claims/total number of claims in the generated model \\
    \\  
    You should ignore formatting or identifier naming differences if the structure and semantics match. \\
    \\
    Input: \\
    Reference Model: \\
    \{reference\_model\} \\
    \\
    Generated Model: \\
    \{generated\_model\} \\
    \\
    Output:
    \end{tcolorbox}
    \caption{Prompt template for SysM-P (precision) evaluation.}
    \label{fig:sysm-p}
\end{figure}

\begin{figure}[htbp]
    \centering
    \begin{tcolorbox}[
        width=\columnwidth,            
        colback=gray!4,                
        colframe=gray!60,              
        sharp corners,                 
        boxrule=0.4pt,                 
        listing only,                  
        listing options={style=sysml}  
    ]
    Your task is to evaluate the recall of a generated system model. You will be given a reference system model and a generated system model. Please perform the following steps: \\
    \\    
    1. List all atomic modeling claims made by the reference system model. Each atomic claim should correspond to a minimal, meaningful modeling element (e.g., the definition of a part, the declaration of an attribute, the use of types, or structural relations like containment or reference). \\
    2. For each atomic claim in the reference model, determine whether it is covered by the generated model (i.e., the generated model contains the same or equivalent element). \\
    3. Summarize the results using the format: Score: number of covered claims/total number of claims in the reference model \\
    \\
    You should ignore formatting or identifier naming differences if the structure and semantics match. \\
    \\
    Input: \\
    Reference Model: \\
    \{reference\_model\} \\
    \\
    Generated Model: \\
    \{generated\_model\} \\
    \\
    Output:
    \end{tcolorbox}
    \caption{Prompt template for SysM-R (recall) evaluation.}
    \label{fig:sysm-r}
\end{figure}

\subsection{F. Studied LLMs.}

\begin{itemize}
    \item \textbf{gpt-4.1-2025-04-14:} OpenAI’s April-2025 flagship model (API only); supports up to \(\approx\)1 M-token context and improves coding and long-context reasoning over GPT-4o. 
    \item \textbf{Claude 3 Opus:} Anthropic’s top-tier Claude-3 variant with a 200 K–1 M token window, optimised for complex reasoning and code. 
    \item \textbf{DeepSeek R1:} 671 B-parameter MoE model (37 B active) released Jan 2025; open-source, math- and reasoning-oriented. 
    \item \textbf{Mistral-7B-Instruct:} Apache-2 licensed 7 B model (Sept 2023) with sliding-window attention for efficient long sequences.  
    \item \textbf{Qwen3-32B:} Alibaba’s bilingual 32 B dense model (2025) featuring enhanced reasoning and instruction following.  
    \item \textbf{gemma-2-9b-it:} Google Gemma 2 instruction-tuned 9 B checkpoint (May 2025), light-weight but strong in English generation.  
    \item \textbf{Llama-3.1-8B-Instruct:} Meta’s June-2025 8 B release from the Llama-3.1 line, instruction-tuned with improved safety.  
    \item \textbf{internlm3-8b-instruct:} Shanghai AI Lab’s third-gen 8 B bilingual model (Jan 2025) licensed under Apache 2.0.  
    \item \textbf{Baichuan2-13B-Chat:} Baichuan AI’s 13 B chat model (2023) trained on 2.6 T tokens, strong on Chinese–English tasks.  
    \item \textbf{ChatGLM3-6B:} Zhipu AI’s 6 B open model (Oct 2024) adding function-calling and code-interpreter skills.  
    \item \textbf{DeepSeek-Coder-V2-Lite:} 16 B MoE code model (Apr 2025) claiming GPT-4-Turbo-level code quality with 130 K context.  
    \item \textbf{Codestral-22B-v0.1:} Mistral’s 22 B code LLM (May 2024) trained on 80 + languages with a 32 K context window. 
    \item \textbf{Phind-CodeLlama-34B-v2:} Phind fine-tune of CodeLlama 34 B achieving 73.8\% pass@1 on HumanEval.  
    \item \textbf{Magicoder-S-CL-7B:} UIUC’s 7 B model built with OSS-Instruct for low-bias, high-quality code instructions.  
    \item \textbf{WizardCoder-15B-V1.0:} WizardLM family 15 B model (Jan 2024) evol-instruct-tuned; strong on HumanEval-Plus. 
    \item \textbf{CodeLlama-13B-Instruct-hf:} Meta’s official 13 B instruct variant for general code synthesis and understanding.  
    \item \textbf{CodeGen2.5-7B-Instruct:} Salesforce’s 7 B instruction-tuned CodeGen 2.5 checkpoint (late 2024) targeting multi-language code generation.  
\end{itemize}

\subsection{G. Experimental Setting for Evaluation.}
Figure~\ref{fig:direct} shows the evaluation prompt.

\begin{figure}[htbp]
    \centering
    \begin{tcolorbox}[
        width=\columnwidth,            
        colback=gray!4,                
        colframe=gray!60,              
        sharp corners,                 
        boxrule=0.4pt,                 
        listing only,                  
        listing options={style=sysml}  
    ]
    You are a senior MBSE engineer. \\
    \\
    Task: \\
    Given the following natural‑language requirements, create an OMG SysML v2 textual model.  \\
    Return only valid SysML v2 code, no explanations or commentary. \\
    \\
    Input Requirement: \\
    \{requirement\} \\
    \\
    Output Model:
    \end{tcolorbox}
    \caption{Prompt template for evaluation.}
    \label{fig:direct}
\end{figure}

\subsection{H. Prompt for Enhancement Strategies.}

Figure~\ref{fig:few-shot}, Figure~\ref{fig:cot} and Figure~\ref{fig:grammar} show the prompts for the three enhancement strategies, seperately. 

\begin{figure}[htbp]
    \centering
    \begin{tcolorbox}[
        width=\columnwidth,            
        colback=gray!4,                
        colframe=gray!60,              
        sharp corners,                 
        boxrule=0.4pt,                 
        listing only,                  
        listing options={style=sysml}  
    ]
    You are a senior MBSE engineer. \\
    \\
    Task: \\
    Given the following natural‑language requirements, create an OMG SysML v2 textual model. \\  
    Return only valid SysML v2 code, no explanations or commentary. \\
    \\
    ––––– FEW‑SHOT EXAMPLES ––––– \\
    \\
    Input Requirements: \\
    {req} \\
    \\
    Output Model: \\
    {design} \\
    ––––– YOUR TURN ––––– \\
    \\
    Input Requirement: \\
    {requirement} \\
    \\
    Output Model:
    \end{tcolorbox}
    \caption{Prompt template for the few-shot strategy.}
    \label{fig:few-shot}
\end{figure}

\begin{figure}[htbp]
    \centering
    \begin{tcolorbox}[
        width=\columnwidth,            
        colback=gray!4,                
        colframe=gray!60,              
        sharp corners,                 
        boxrule=0.4pt,                 
        listing only,                  
        listing options={style=sysml}  
    ]
    You are a senior MBSE engineer. \\
    \\
    Task: \\
    1. Think step‑by‑step in a hidden scratchpad (not shown to user) \\
    \hspace*{0.8em} - Extract key functional/non‑functional information. \\
    \hspace*{0.8em} - Map them to various grammars in the SysML v2 textual grammar. \\
    2. After thinking, output only valid SysML v2 textual code—no explanations, no scratchpad. \\
    \\
    Input Requirement: \\
    {requirement} \\
    \\
    Output Model:
    \end{tcolorbox}
    \caption{Prompt template for the CoT strategy.}
    \label{fig:cot}
\end{figure}

\begin{figure}[htbp]
    \centering
    \begin{tcolorbox}[
        width=\columnwidth,            
        colback=gray!4,                
        colframe=gray!60,              
        sharp corners,                 
        boxrule=0.4pt,                 
        listing only,                  
        listing options={style=sysml}  
    ]
    You are a senior MBSE engineer. \\
    \\
    Task: \\
    Given the following natural‑language requirements, create an OMG SysML v2 textual model.  \\
    Your output must conform to the BNF grammar below (subset of SysML v2). \\ 
    Return only valid SysML v2 code, no explanations or commentary. \\
    \\
    ––––– SysML v2 BNF (subset) –––––\\
    \{bnf\_grammar\}
    ––––––––––––––––––––––––––––––––––\\
    \\
    Input Requirement: \\
    \{requirement\}
    \\
    Output System Model: 
    \end{tcolorbox}
    \caption{Prompt template for the grammar strategy.}
    \label{fig:grammar}
\end{figure}

\subsection{I. Error Cases for Claude 3.}
Figure~\ref{fig:error_analysis} shows the case to explain the error of Claude with CoT prompting strategies.

\begin{figure}[htbp]
    \centering
    \begin{subfigure}[t]{\columnwidth}
        \centering
        \begin{tcolorbox}[
            width=\columnwidth,
            colback=gray!4,
            colframe=gray!60,
            sharp corners,
            boxrule=0.4pt,
            listing only,
            listing options={style=sysml}
        ]
        package VehicleStructure \{ \\
        \hspace*{0.8em} abstract part def VehicleComponent; \\
        \hspace*{0.8em} part def Engine :\> VehicleComponent; \\
        \hspace*{0.8em} part def Transmission :\> VehicleComponent; \\
        \hspace*{0.8em} part def Wheel :\> VehicleComponent; \\
        \hspace*{0.8em} abstract part def Vehicle \{ \\
        \hspace*{1.6em} part engine : Engine[1]; \\
        \hspace*{1.6em} part transmission : Transmission[1]; \\
        \hspace*{1.6em} part wheels : Wheel[4]; \\
        \hspace*{0.8em} \} \\
        \} 
        \end{tcolorbox}
        \caption{The Generated System Model by Claude 3}
        \label{fig:issue_1}
    \end{subfigure}

    \begin{subfigure}[t]{\columnwidth}
        \centering
        \begin{tcolorbox}[
            width=\columnwidth,
            colback=gray!4,
            colframe=gray!60,
            sharp corners,
            boxrule=0.4pt,
            listing only,
            listing options={style=sysml}
        ]
        package 'VehicleDefinition' { \\
	part def Vehicle { \\
		part parts : VehiclePart[*]; \\
		part eng : Engine subsets parts; \\
		part trans : Transmission subsets parts; \\
		part wheels : Wheel[4] :\> parts; \\
	}
	abstract part def VehiclePart; \\
	part def Engine :\> VehiclePart; \\
	part def Transmission :> VehiclePart; \\
	part def Wheel :\> VehiclePart; 
        }
        \end{tcolorbox}
        \caption{The Ground Truth System Model}
        \label{fig:issue_2}
    \end{subfigure}

    \caption{Illustrations of the error of Claude with CoT prompting strategies. }
    \label{fig:error_analysis}
\end{figure}

\subsection{J. Ground Truth of Case Study.}
Figure~\ref{fig:casestudy} shows the ground truth of the case study.

\begin{figure}[htbp]
    \centering
    \begin{tcolorbox}[
        width=\columnwidth,            
        colback=gray!4,                
        colframe=gray!60,              
        sharp corners,                 
        boxrule=0.4pt,                 
        listing only,                  
        listing options={style=sysml}  
    ]
    package 'TrafficLightDefinition' \{ \\
    \hspace*{0.8em} private import ScalarValues::Real; \\
    \hspace*{0.8em} enum def TrafficLightColor \{ \\
    \hspace*{1.6em} enum green; \\
    \hspace*{1.6em} enum yellow; \\
    \hspace*{1.6em} enum red; \\
    \hspace*{0.8em} \} \\
    \hspace*{0.8em} part def TrafficLight \{ \\
    \hspace*{1.6em} attribute currentColor : TrafficLightColor; \\
    \hspace*{0.8em} \} \\
    \hspace*{0.8em} part def TrafficLightGo specializes TrafficLight \{ \\
    \hspace*{1.6em} attribute redefines currentColor =  TrafficLightColor::green; \\
    \hspace*{0.8em} \} \\
    \}
    \end{tcolorbox}
    \caption{Groundtruth of the case study.}
    \label{fig:casestudy}
\end{figure}

\subsection{K. More Analysis on Key Grammar.}
Table~\ref{tab:metrics-part1} and Table~\ref{tab:metrics-part2} show the performance of Qwen 3-32B on each key grammar.

\begin{table*}[]
    \captionsetup{justification=centering} 
    \centering
    \caption{Per-category evaluation scores (Part 1)}
    \label{tab:metrics-part1}
    \resizebox{\textwidth}{!}{%
    \begin{tabular}{lrrrrr}
        \toprule
        \textbf{Category} & \textbf{BLEU} & \textbf{ROUGE} & \textbf{BERTScore} & \textbf{SysM~P} & \textbf{SysM~R}\\
        \midrule
        Attribute                    & 0.058 & 0.438 & 0.723 & 0.700 & 0.455\\
        Generalization               & 0.015 & 0.443 & 0.610 & 1.000 & 1.000\\
        Subsetting                   & 0.066 & 0.504 & 0.695 & 0.200 & 0.778\\
        Redefinition                 & 0.061 & 0.616 & 0.736 & 0.769 & 0.923\\
        Enumeration                  & 0.027 & 0.478 & 0.683 & 1.000 & 0.813\\
        Part                         & 0.014 & 0.441 & 0.668 & 0.702 & 0.675\\
        Item                         & 0.019 & 0.368 & 0.675 & 0.625 & 0.100\\
        Connection                   & 0.036 & 0.481 & 0.708 & 0.778 & 0.818\\
        Port                         & 0.034 & 0.431 & 0.699 & 0.671 & 0.331\\
        Function-based Behavior      & 0.034 & 0.477 & 0.654 & 0.763 & 0.572\\
        Interface                    & 0.022 & 0.458 & 0.672 & 0.250 & 0.571\\
        State-based Behavior         & 0.004 & 0.405 & 0.563 & 0.754 & 0.391\\
        Individual \& Snapshot       & 0.002 & 0.293 & 0.625 & 0.662 & 0.857\\
        Variant Configuration        & 0.044 & 0.457 & 0.723 & 0.951 & 0.905\\
        Requirement                  & 0.013 & 0.408 & 0.659 & 0.328 & 0.453\\
        Verification                 & 0.012 & 0.386 & 0.643 & 0.606 & 0.591\\
        Analysis and Trade           & 0.021 & 0.413 & 0.681 & 0.643 & 0.643\\
        View and Viewpoint           & 0.023 & 0.467 & 0.699 & 0.875 & 0.688\\
        Dependency                   & 0.011 & 0.479 & 0.627 & 0.826 & 0.476\\
        Model Constrainment          & 0.017 & 0.493 & 0.651 & 0.589 & 0.968\\
        Binding Connector            & 0.009 & 0.363 & 0.669 & 0.188 & 0.283\\
        Language Extension           & 0.015 & 0.432 & 0.654 & 0.517 & 0.363\\
        Expression                   & 0.024 & 0.430 & 0.669 & 0.697 & 0.670\\
        \bottomrule
    \end{tabular}
    }
\end{table*}

\begin{table*}[!t]
    \centering
    \caption{Per-category evaluation scores (Part 2)}
    \label{tab:metrics-part2}
    \resizebox{\textwidth}{!}{%
    \begin{tabular}{lrrrrr}
        \toprule
        \textbf{Category} & \textbf{BLEU} & \textbf{ROUGE} & \textbf{BERTScore} & \textbf{SysM~P} & \textbf{SysM~R}\\
        \midrule
        Sequence Modeling            & 0.017 & 0.418 & 0.615 & 0.652 & 0.407\\
        Flow Connection              & 0.033 & 0.496 & 0.732 & 0.566 & 0.629\\
        Action Definition            & 0.011 & 0.399 & 0.565 & 0.542 & 0.513\\
        Action                       & 0.060 & 0.464 & 0.704 & 0.938 & 0.600\\
        Conditional Succession       & 0.023 & 0.393 & 0.643 & 0.583 & 0.495\\
        Control Structure            & 0.023 & 0.448 & 0.682 & 0.590 & 0.308\\
        Assignment Action            & 0.016 & 0.421 & 0.622 & 0.917 & 0.657\\
        Message                      & 0.056 & 0.492 & 0.710 & 0.664 & 0.828\\
        Opaque Action                & 0.041 & 0.321 & 0.700 & 0.714 & 0.200\\
        State Definition             & 0.012 & 0.510 & 0.619 & 1.000 & 0.841\\
        State                        & 0.019 & 0.509 & 0.681 & 0.752 & 0.628\\
        Transition                   & 0.020 & 0.473 & 0.671 & 0.784 & 0.623\\
        Occurrence                   & 0.021 & 0.450 & 0.663 & 0.692 & 0.609\\
        Individual                   & 0.042 & 0.504 & 0.686 & 0.623 & 0.721\\
        Calculation                  & 0.007 & 0.421 & 0.592 & 0.696 & 0.611\\
        Constraint                   & 0.032 & 0.464 & 0.678 & 0.563 & 0.446\\
        Analysis                     & 0.011 & 0.404 & 0.613 & 0.512 & 0.638\\
        Use Case                     & 0.014 & 0.368 & 0.573 & 0.806 & 0.472\\
        Variability                  & 0.044 & 0.529 & 0.668 & 0.719 & 0.755\\
        Functional Allocation        & 0.051 & 0.558 & 0.760 & 0.563 & 0.418\\
        Metadata                     & 0.015 & 0.317 & 0.612 & 0.895 & 0.699\\
        Filtering                    & 0.033 & 0.529 & 0.728 & 0.593 & 0.825\\
        View                         & 0.010 & 0.428 & 0.539 & 0.221 & 0.241\\
        Package                      & 0.048 & 0.371 & 0.570 & 0.633 & 0.282\\
        \bottomrule
    \end{tabular}}
\end{table*}

\end{document}